\documentclass{ws-ijmpd}
\usepackage[super]{cite}
\usepackage{xcolor}
\usepackage[verbose,hypertexnames=false]{hyperref}
\hypersetup{colorlinks=false,allbordercolors=blue,pdfborderstyle={/S/U/W 1}}

\begin{document}

\markboth{Owais Ullah Faiz, Mushahid Hussain, and Shashank Shalgar}
{Can we ignore the time dependence in matter neutrino resonance?}

%
\catchline{}{}{}{}{}
%

\title{Can we ignore the time dependence in matter neutrino resonance?}

\author{Owais Ullah Faiz}

\address{Department of Physics and Astronomy,\\
National University of Sciences and Technology,\\
Islamabad, Pakistan.\\
owise.faiz@gmail.com}

\author{Mushahid Hussain}

\address{
Department of Physics and Astronomy,\\
National University of Sciences and Technology,\\
Islamabad, Pakistan.\\
mushahidhussain0073@gmail.com
}

\author{Shashank Shalgar}

\address{
Niels Bohr International Academy and DARK, \\
Niels Bohr Institute, \\
University of Copenhagen, \\
Blegdamsvej 17, 2100, Copenhagen, Denmark.\\
shashank.shalgar@nbi.ku.dk
		}

\maketitle


\begin{abstract}
In the vicinity of neutron star mergers (NSMs), it is possible for the neutrino self-interaction potential to cancel with the matter potential leading to matter neutrino resonance (MNR). 
MNR is one of the most interesting mechanisms by which neutrino flavor evolution can occur in dense astrophysical environments. 
Previous studies have typically assumed that the neutrino flavor field evolves to a steady state—a simplification also used in other self-interaction models such as the neutrino-bulb model. Here, we perform reproducible calculations of MNR using both time-independent and time-dependent formalisms and show that they yield qualitatively different flavor survival probabilities.
The time-independent approach produces unstable steady-state solutions that differ fundamentally from the dynamical behavior captured in time-dependent simulations. These results demonstrate that the steady-state assumption is generally invalid, and physical interpretations based on time-independent calculations of dense neutrino systems require re-evaluation.
\end{abstract}


\section{Introduction}	

In dense astrophysical environments, neutrino flavor evolution acquires a new degree of phenomenological richness due to neutrino self-interactions -- a potential experienced by neutrinos due to other neutrinos in the medium~\cite{Pantaleone:1992eq}. The evolution of neutrino flavor in such environments exhibits a collective evolution of neutrino flavor, that is, the evolution of the various momentum modes is correlated. The phenomenon of collective neutrino oscillations is non-linear in nature, and most of the progress in the field and our understanding is based on numerical simulations of the phenomenon that are extremely challenging~\cite{Duan2006,Duan2006b,Duan:2010bg}. In certain cases, known as fast flavor conversions, the time scale of flavor evolution can be extremely short, posing even more challenges from a numerical point of view~\cite{Sawyer:2005jk, Sawyer:2015dsa}.

Early investigations of collective neutrino oscillations primarily focused on the neutrino-bulb model -- a spherically symmetric toy model designed to capture the essential physics of flavor evolution in the interior of core-collapse supernovae (CCSNe). Although this model provided valuable insights, it incorporated several idealized assumptions~\cite{Duan2006,Duan2006b,Fogli:2007bk,Duan:2010bg}. In particular, it assumes that all neutrinos decouple at a single radius and that their emission is perfectly spherically symmetric. More importantly, the formalism presupposes that the flavor field reaches a steady state after a sufficiently long evolution time. This assumption, though not initially recognized as problematic, was effectively built into the theoretical framework of the bulb model.

Apart from supernovae, another environment where neutrino self-interactions become important is in the regions surrounding neutron star mergers (NSMs)~\cite{Tamborra:2020cul,Johns:2025mlm}. A crucial difference between neutrinos emitted from supernovae and neutron star mergers is that in neutron star mergers, there is an excess of electron antineutrinos over electron neutrinos, unlike supernovae, where the situation is opposite. This leads to an interesting phenomenon of matter neutrino resonance (MNR), which results from the matter potential canceling the self-interaction potential, giving rise to resonant neutrino flavor conversion.~\cite{Malkus:2014iqa}. The phenomenon was originally described in the single-angle approximations. In the single-angle approximation the dependence of the Hamiltonian on the angle is ignored. The effects of relaxing the single-angle approximation have also been studied~\cite{Shalgar:2017pzd,Vlasenko:2018irq,Padilla-Gay:2024wyo}.

To date, however, all multi-angle simulations of MNR have employed a formalism that assumes the flavor field evolves toward a steady state. In contrast, recent work on collective oscillations in supernovae indicates that the flavor field does not generally reach such a steady configuration~\cite{Shalgar:2025ekb}. This raises an important question: does the MNR phenomenon—described by similar equations of motion but under different initial conditions—also exhibit non-steady behavior? In this study, we find that although the MNR system appears to evolve toward a steady state, the steady-state solutions obtained from the time-independent formalism are intrinsically unstable. Even small perturbations drive the system away from these solutions, leading to dynamics that are fundamentally different from those captured by the full time-dependent treatment.

In this paper, we use a simple multi-angle model of MNR to test the validity of the assumption of time-independent formalism in the context of multi-angle calculations of MNR~\cite{Shalgar:2017pzd,Vlasenko:2018irq,Padilla-Gay:2024wyo}. We perform multi-angle simulations of MNR both with and without the time-independent assumption, and find that the two formalisms yield qualitatively different results. This implies that the widely used time-independent formalism is generally not valid. 
Moreover, the implications of this study extend beyond the MNR phenomenon: the same formalism has been widely applied to neutrino flavor evolution in supernovae through the neutrino-bulb model. Consequently, physical insights derived from such steady-state analyses warrant re-examination. Revisiting these foundational assumptions is essential for developing a more accurate understanding of neutrino flavor dynamics in realistic astrophysical environments.

This paper is organized as follows: In Sec.~\ref{formalism}, we discuss the phenomenon of matter neutrino resonance with an emphasis on the distinction between time-dependent and time-independent formalism. In Sec.~\ref{numerical-setup} we explain the numerical setup used in this paper, and in Sec.~\ref{numerical-results} we present the numerical results. In Sec.~\ref{conclusions}, we discuss the implications of the numerical results and conclude. In addition, we present convergence tests for our results in ~\ref{AppendixA}\footnote{The code used for the results in this paper can be found at \url{https://github.com/shashankshalgar/MNRpaper2025}}.

\section{Formalism: Time-Dependent and Time-Independent Equations for MNR}
{\label{formalism}}
The numerical evolution of neutrinos experiencing MNR is difficult to model numerically due to the complex geometry associated with the problem. Attempts to investigate MNR using realistic potentials are limited to the single-angle approximation, due to the complicated geometry~\cite{Zhu:2016mwa}. If the single-angle approximation is relaxed, then unrealistic symmetries have to be imposed to understand the phenomenon of MNR. One such attempt invoked an infinite disk that emits neutrinos along a particular opening angle~\cite{Shalgar:2017pzd}, and another attempt has assumed spherical symmetry for the MNR system~\cite{Vlasenko:2018irq}. Although neither approach is clearly justified, we use the former on account of its simplicity. The goal of this paper is not to model the neutrino transport in a system of neutron star mergers but to investigate the assumption of time-independence of the solution.  In particular, we investigate the justification used in previous papers that calculates neutrino flavor evolution in the presence of MNR. In what follows, we first present the time-dependent formalism and then derive the time-independent equations from it.

\subsection{Time-Dependent Formalism}

Throughout this paper, we restrict our analysis to the two-flavor approximation, in which the flavor states of neutrinos and antineutrinos are represented in terms of $2 \times 2$ density matrices. The diagonal components of the density matrix denote the occupation numbers, whereas the off-diagonal components contain information about coherence between the flavor eigenstates. Following the approach used in \cite{Shalgar:2017pzd}, we perform the simulation in a narrow range of angles,
\begin{eqnarray}
\cos \theta \in [\cos \theta_{\mathrm{min}}, \cos \theta_{\mathrm{max}}] \ .
\end{eqnarray}
Throughout this paper, we use a range for $\theta$ that is centered at $45^{\circ}$ but various values for the range, $\Delta \theta$ defined as $|\theta_{\mathrm{max}} - \theta_{\mathrm{min}}|$. This allows us to clearly see the effect of using multi-angle formalism by changing the value of $\Delta \theta$ because the results are the same as in the single-angle case in the limit of $\Delta \theta \rightarrow 0$. To simplify the problem from a numerical point of view, we ignore the collision term and only consider the advective term along with the flavor evolution as shown below:
\begin{eqnarray}
		\label{eom1}
		i \left(\frac{\partial}{\partial t} + \cos\theta \frac{\partial}{\partial r}\right) \rho(\cos\theta,r,t) &=& [H(\cos\theta,r,t),\rho(\cos\theta,r,t)]\\
		\label{eom2}
		i \left(\frac{\partial}{\partial t} + \cos\theta \frac{\partial}{\partial r}\right) \bar{\rho}(\cos\theta,r,t) &=& [\bar{H}(\cos\theta,r,t),\bar{\rho}(\cos\theta,r,t)] \ . 
\end{eqnarray}
The total derivative in the parentheses on the left-hand side of the equations is used, assuming infinite extent of the source. For a finite source, the total derivative would include a derivative with respect to the angular variables, which we have not included for simplicity. It should also be noted that we have neglected the energy dependence of the density matrices, as it is not relevant in the context of MNR. The Hamiltonian for neutrinos as well as antineutrinos consists of three components corresponding to the vacuum term, the matter term, and the self-interaction term,
\begin{eqnarray}
		H(\cos\theta,r,t) &=& H_{\mathrm{vac}} + H_{\mathrm{mat}} + H_{\nu\nu}\\  
		\bar{H}(\cos\theta,r,t) &=& -H_{\mathrm{vac}} + H_{\mathrm{mat}} + H_{\nu\nu} \ ,
\end{eqnarray}
where,
\begin{eqnarray}
		\label{Hvac}
		H_{\mathrm{vac}} &=& \frac{\omega_{\mathrm{vac}}}{2} 
		\begin{pmatrix}
				-\cos 2\vartheta_{\mathrm{V}} & \sin 2 \vartheta_{\mathrm{V}} \\
				\sin 2\vartheta_{\mathrm{V}} & \cos 2\vartheta_{\mathrm{V}}
		\end{pmatrix}\ ,\\
		\label{Hmat}
		H_{\mathrm{mat}}(r) &=& \begin{pmatrix}
				\sqrt{2}G_{\mathrm{F}}n_{e}(r) & 0 \\
                0 & 0
        \end{pmatrix}\ ,\\
		\label{Hself}
		H_{\nu\nu}(\cos\theta,r,t) &=& \mu(r) \int_{\cos\theta_{\mathrm{min}}}^{\cos\theta_{\mathrm{max}}} (\rho(\cos\theta^{\prime},r,t) - \bar{\rho}(\cos\theta^{\prime},r,t)) \nonumber \\
		& \times & (1-\cos\theta\cos\theta^{\prime})
		d\cos\theta^{\prime} \ .
\end{eqnarray}

Throughout this paper we use $\omega_{\mathrm{vac}} = 1$ km$^{-1}$ and $-1$ km$^{-1}$ corresponding to normal and inverted mass ordering, respectively. The value of the mixing angle is set at $\vartheta_{\mathrm{V}} = 0.15$ radians. 
In the definition of the Hamiltonians, we use $G_{\mathrm{F}}$ to denote the Fermi constant and $n_{e}(r)$ to denote the number density of electrons, while $n_{\nu}$ denotes the number density of neutrinos (not neutrinos and antineutrinos). Throughout the rest of the paper, we use $\mu(r)$ to denote the combination $\sqrt{2}G_{\mathrm{F}}n_{\nu}(r)$. The matter Hamiltonian is dependent only on the number density of electrons because we ignore the potential due to neutral current interaction, as it is the same for neutrinos of all flavors, and hence, only adds an unobservable overall phase to the neutrino wavefunction. It is possible to define potentials associated with the matter and the self-interaction Hamiltonians as $V_{e}(r) = \sqrt{2}G_{\mathrm{F}}n_{e}(r)$ and $V_{\nu\nu}(\cos\theta,r,t) = H_{\nu\nu}^{ee}(\cos\theta,r,t) - H_{\nu\nu}^{xx}(\cos\theta,r,t)$, respectively, where $H_{\nu\nu}^{ee}$ and $H_{\nu\nu}^{xx}$ are the diagonal components of the self-interaction Hamiltonian. Unlike the matter potential, the self-interaction potential is an angle-dependent quantity; however, this dependence becomes less important as $\Delta \theta$ decreases.  

The NSM is a neutron-rich system that emits more electron antineutrinos than electron neutrinos, which means that the neutrino self-interaction potential is negative, unlike that in the context of core-collapse supernovae. This suggests that the self-interaction potential could cancel the matter potential, thereby enabling resonant flavor evolution. This possibility was first pointed out in the context of single-angle approximation\cite{Malkus:2014iqa}. In the single-angle approximation, the cancellation between the two potentials is complete due to both potentials being independent of $\cos\theta$. However, it was soon realized that relaxing the single-angle approximation leads to a weaker resonance as the cancellation of the potentials can only occur for a neutrino with a given $\cos\theta$ at any given time~\cite{Shalgar:2017pzd, Vlasenko:2018irq}. All the multi-angle calculations were performed using the time-independent formalism, which can be derived by assuming that Eqs.~\ref{eom1} and \ref{eom2} result in a steady state. 

\subsection{Time-Independent Formalism}
In the time-independent formalism, temporal derivatives in Eqs.~\ref{eom1} and~\ref{eom2} are neglected, effectively assuming that the flavor field evolves to a steady-state configuration. The equations of motion then reduce to,
\begin{eqnarray}
		\label{timeindepeom1}
i\cos\theta \frac{\partial}{\partial r} \rho(\cos\theta,r) &=& [H(\cos\theta,r),\rho(\cos\theta,r)]\\
		\label{timeindepeom2}
i\cos\theta \frac{\partial}{\partial r} \bar{\rho}(\cos\theta,r) &=& [\bar{H}(\cos\theta,r),\bar{\rho}(\cos\theta,r)] \ .
\end{eqnarray}
This is an initial value problem with $r$ as the independent variable, as long as $\cos\theta$ is greater than zero for all neutrinos or $\cos\theta$ is less than zero for all neutrinos. This condition on $\cos\theta$ is valid when neutrinos are emitted from a surface, as in the neutrino-bulb model, but not in more realistic models with a finite decoupling region. 

\section{Numerical set-up}
\label{numerical-setup}
In this section, we compare results obtained using the time-dependent and time-independent formalisms for the same physical system.
The formalism used for the time-independent solution is identical to the one used in earlier papers on this topic~\cite{Shalgar:2017pzd, Vlasenko:2018irq}. 
The time-independent approach follows the methods employed in previous MNR studies~\cite{Shalgar:2017pzd,Vlasenko:2018irq}, as well as in the simplified supernova neutrino-bulb model~\cite{Duan2006,Duan2006b,Fogli:2007bk,Duan:2010bg}. To our knowledge, however, no study to date has implemented the fully time-dependent formalism in the context of MNR. 

\subsection{Time-independent case}

We solve the MNR problem in the time-independent formalism as described in Eqs.~\ref{timeindepeom1} and \ref{timeindepeom2} by discretizing the density matrix uniformly along the range of $\cos\theta$. 
We initialize the density matrices such that initially there is no $\cos\theta$ dependence and normalized such that at $r=r_{\mathrm{min}}$ such that at $r=r_{\mathrm{min}}$,
\begin{eqnarray}
		\int_{\cos\theta_{\mathrm{min}}}^{\cos\theta_{\mathrm{max}}} \rho_{ee}(\cos\theta,r_{\mathrm{min}}) d \cos\theta &=& 1 \\
		\int_{\cos\theta_{\mathrm{min}}}^{\cos\theta_{\mathrm{max}}} \bar{\rho}_{ee}(\cos\theta,r_{\mathrm{min}}) d \cos\theta &=& 4/3 \\
		\int_{\cos\theta_{\mathrm{min}}}^{\cos\theta_{\mathrm{max}}} \rho_{xx}(\cos\theta,r_{\mathrm{min}}) d \cos\theta &=& 0 \\
		\int_{\cos\theta_{\mathrm{min}}}^{\cos\theta_{\mathrm{max}}} \bar{\rho}_{xx}(\cos\theta,r_{\mathrm{min}}) d \cos\theta &=& 0 \ .
\end{eqnarray}
Throughout this paper, we use a matter potential that is independent of the radius and a self-interaction potential that falls with the radius as described below:
\begin{eqnarray}
		V_{e}(r) &=& 1000 \mathrm{\ km}^{-1} \\
		\mu(r) &=& \frac{2 \times 10^{7}}{r^{3}} \mathrm{\ km}^{-1} \ ,
\end{eqnarray}
where $\mu(r)$ is used as a radially dependent coefficient in the self-interaction Hamiltonian as follows,
\begin{eqnarray}
		\label{Hnunu:timeindep}
		H_{\nu\nu}(r) = \mu(r) \int_{\cos\theta_{\mathrm{min}}}^{\cos\theta_{\mathrm{max}}} (\rho(\cos\theta^{\prime},r) - \bar{\rho}(\cos\theta^{\prime},r)) \times (1-\cos\theta \cos\theta^{\prime}) d \cos\theta^{\prime} \ .
\end{eqnarray}
It should be noted that in the time-independent formalism, the radial dependence is equivalent to temporal dependence and hence the Hamiltonian above is the same as the one in Eq.~\ref{Hself} apart from the missing time dependence.

Although a matter potential that does not fall with the radius is unrealistic, we would like to emphasize that our goal is not to simulate a realistic MNR scenario but to demonstrate the incorrectness of the time-independent formalism that has been widely used. The same holds for the radial dependence of the strength of the self-interaction potential.

With the given initial conditions and the Hamiltonian, it is possible to solve the time-independent equations of motion, Eqs.~\ref{timeindepeom1} and \ref{timeindepeom2} as an initial value problem for the density matrix for each $\cos\theta$ bin. It should be noted that the density matrices for $\cos\theta$ bins are coupled due to the self-interaction Hamiltonian. 

In this paper, we solve the time-independent equations of motion using \texttt{DifferentialEquations.jl} in Julia. We use the VCABM3 solver with an adaptive step size with absolute and relative tolerances of $10^{-9}$ each.  

\subsection{Time-dependent case}

For the time-dependent case, there is an additional degree of freedom as $r$ and $t$ are not interchangeable as seen in Eqs.~\ref{eom1} and \ref{eom2}. The density matrices are initialized to be uniformly distributed along $\cos\theta$ as well as $r$. We normalize the density matrices to have the same initial values as in the previous subsection; however, the density matrices are initialized to those values for each spatial bin.
\begin{eqnarray}
		\int_{\cos\theta_{\mathrm{min}}}^{\cos\theta_{\mathrm{max}}} \rho_{ee}(\cos\theta,r,t_{\mathrm{min}}) d \cos\theta &=& 1 \\
        \int_{\cos\theta_{\mathrm{min}}}^{\cos\theta_{\mathrm{max}}} \bar{\rho}_{ee}(\cos\theta,r,t_{\mathrm{min}}) d \cos\theta &=& 4/3 \\
        \int_{\cos\theta_{\mathrm{min}}}^{\cos\theta_{\mathrm{max}}} \rho_{xx}(\cos\theta,r,t_{\mathrm{min}}) d \cos\theta &=& 0 \\
        \int_{\cos\theta_{\mathrm{min}}}^{\cos\theta_{\mathrm{max}}} \bar{\rho}_{xx}(\cos\theta,r,t_{\mathrm{min}}) d \cos\theta &=& 0 \ .
\end{eqnarray}
As $r$ and $t$ are not interchangeable, we initialize the density matrices at the initial time, and the initialization is the same across all the radial bins.
We use the same radial profile for $\mu(r)$ as used in the previous subsection in Eq.~\ref{Hself}. 

Unlike in the time-independent case, the advective term that involves a derivative needs to be calculated using a numerical scheme. We use the central difference method to calculate the spatial derivative, while the time evolution is performed using the VCABM3 solver with an adaptive step size, as before. The size of the spatial grid used and the justification for that are given in Sec .~\ref {numerical-results} and \ref{AppendixA}. We evolve the system of equations until the survival probability stops evolving. 

The complete implementation, including input parameters and analysis scripts, is publicly available at \url{https://github.com/shashankshalgar/MNRpaper2025}.

\section{Numerical results}
\label{numerical-results}

In this section, we compare the results obtained using the same system using time-dependent and time-independent formalisms. We present the results for three different opening angles: $\Delta \theta = 1^{\circ}, 2^{\circ}$, and $4^{\circ}$. 
For all cases, we fix $r_{\mathrm{min}} = 10~\mathrm{km}$ and $r_{\mathrm{max}} = 60~\mathrm{km}$. In the time-independent formalism, the radial coordinate effectively plays the role of time in determining the flavor evolution. Because this system develops a cascade to small angular scales—where mode coupling generates fine angular structure—a large number of $\cos\theta$ bins is required for convergence, especially for $\Delta\theta = 4^{\circ}$. All results shown here use a sufficient number of angular bins to ensure numerical convergence, with examples provided in Appendix~\ref{AppendixA}.

In contrast, the time-dependent formalism does not exhibit such angular-scale cascades. Instead, the key factor controlling convergence is spatial resolution. We find that a spatial grid spacing of $0.025~\mathrm{km}$ yields converged survival probabilities in all cases, as demonstrated in Appendix~\ref{AppendixA}.

\begin{figure}
		\includegraphics[width=0.99\textwidth]{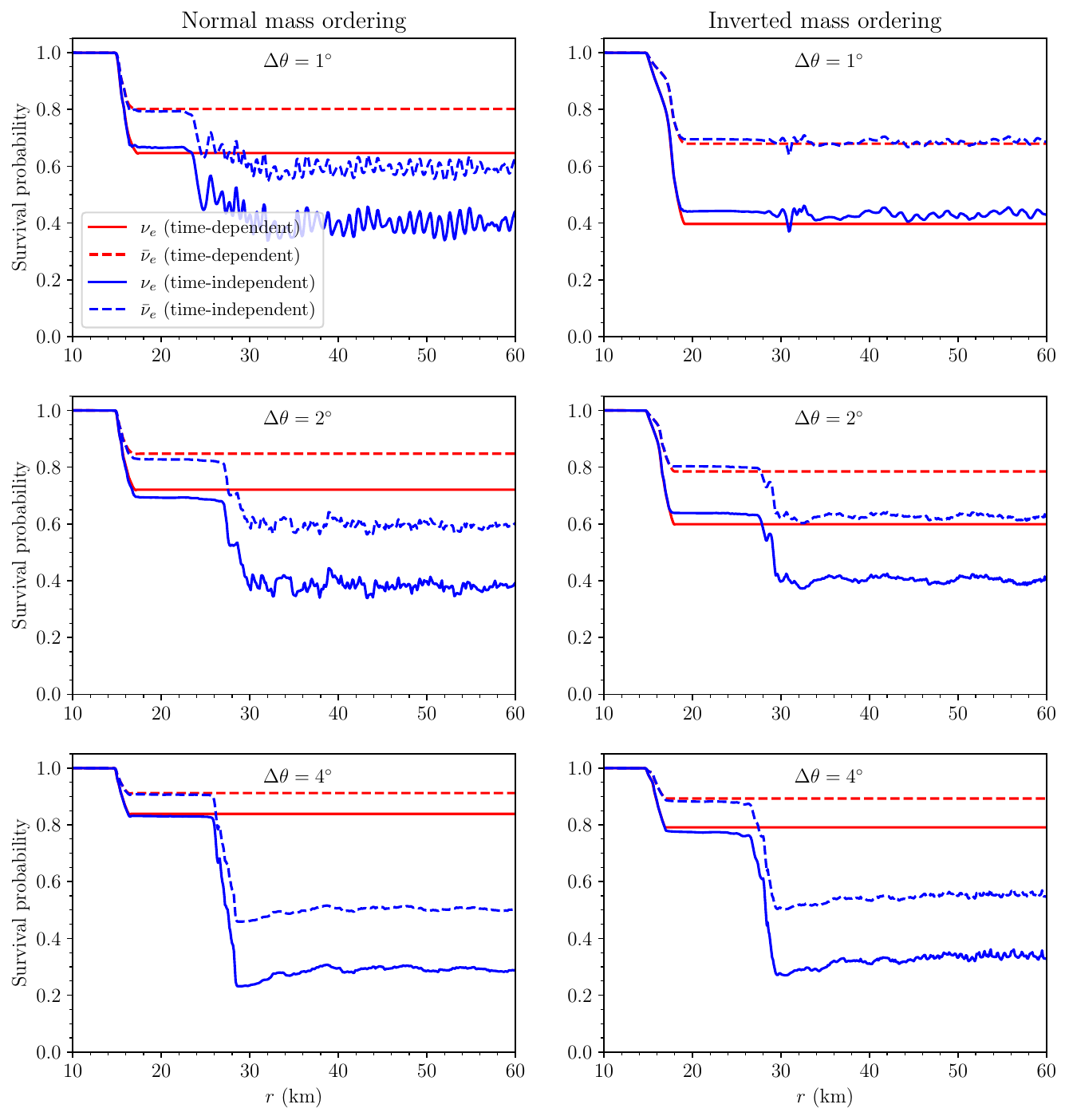}
		\caption{{\it Left: }Comparison between neutrino flavor evolution for normal mass odering in identical systems obtained using time-dependent formalism (red) and time-independent formalism (blue) The top, middle, and bottom panels are for $\Delta \theta = 1^{\circ}, 2^{\circ}$, and $4^{\circ}$, respectively. {\it Right: } Same as that in the left column but for inverted mass ordering.}
		\label{maincomp}
\end{figure}

\begin{figure}
        \includegraphics[width=0.49\textwidth]{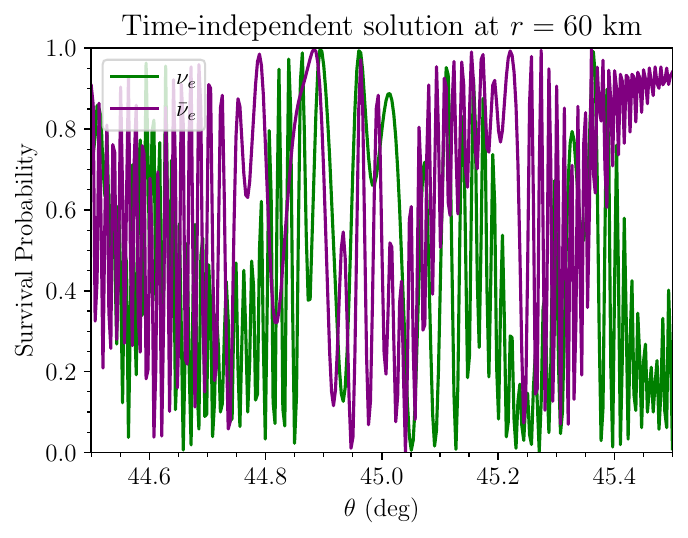}
		\includegraphics[width=0.49\textwidth]{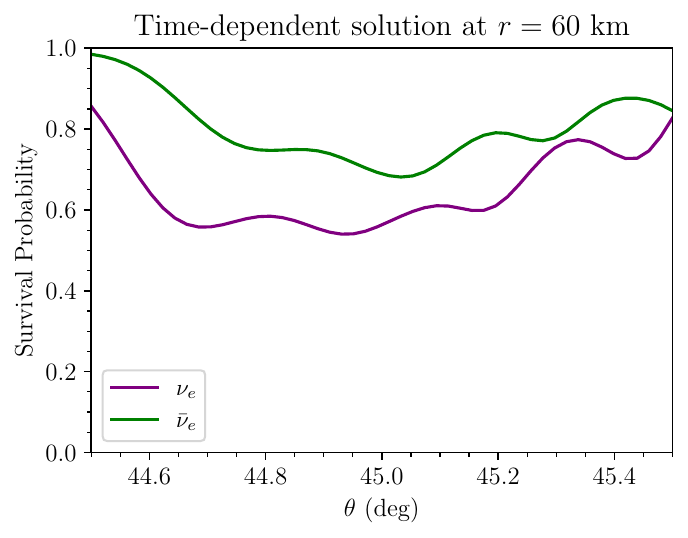}
		\caption{{\it Left: } Angular distribution of electron neutrinos (green) and electron antineutrinos (purple) at 60 km obtained using time-independent formalism and assuming normal mass ordering. One can see that the angular distribution has significant structure, and hence, a large number of angular bins is required. The results in this panel have been obtained with 100 angular bins, which is sufficient in this case, as seen in~\ref{AppendixA}. {\it Right:} Angular distribution of electron neutrinos (green) and electron antineutrinos (purple) at 60 km obtained using time-dependent formalism. The angular distribution does not have too much structure, and hence, the number of angular bins required is not large. The results in this panel have been obtained using 50 angular bins, which is more than what is necessary for convergence.}
        \label{angdist}
\end{figure}

Figure~\ref{maincomp} shows the survival probabilities of neutrinos and antineutrinos for both formalisms. The two approaches produce clearly distinct results: for all three values of $\Delta\theta$, the time-independent formalism consistently overestimates the degree of flavor conversion. Moreover, in the time-independent case, the survival probability remains approximately constant for several kilometers beyond the resonance before changing rapidly. This behavior reflects the continued cascade to finer angular scales. The left panel of Fig.~\ref{angdist} illustrates this cascade through the development of angular structure in the neutrino and antineutrino distributions.

For the time-dependent case, the angular structure ceases to develop after the first drop in the survival probability; both the angular distribution and survival probability remain constant thereafter, as shown in the right panel of Fig.~\ref{angdist}. The number of $\cos\theta$ bins required for convergence is much smaller in the case of time-dependent formalism for this reason. A more detailed study of the convergence of the results presented in this paper can be found in~\ref{AppendixA}. 

\begin{figure}
        \includegraphics[width=0.99\textwidth]{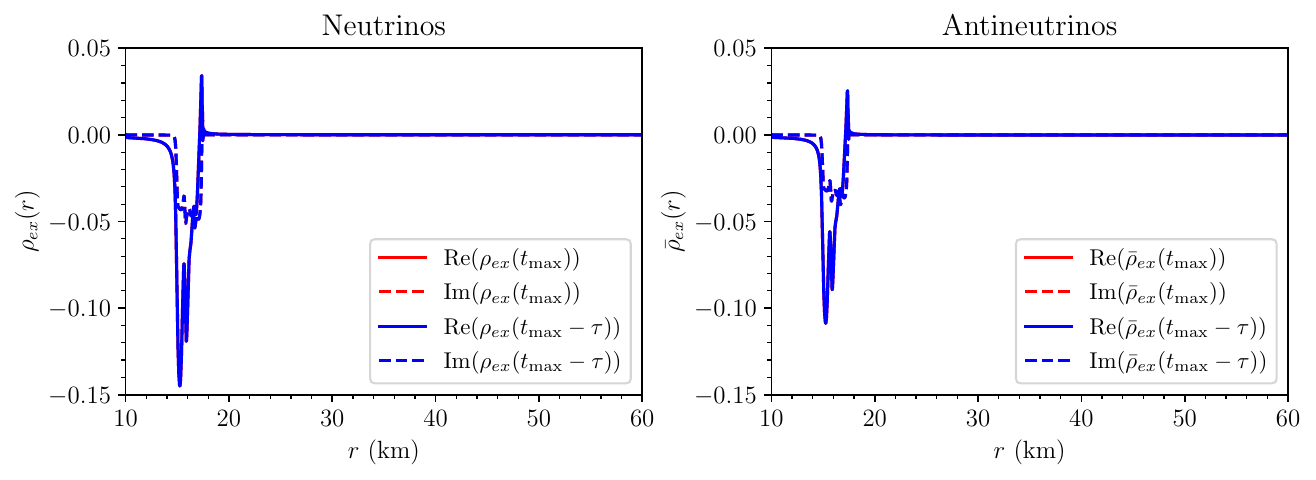}
		\caption{{\it Left:} Real (solid) and imaginary (dashed) parts of the off-diagonal component of the density matrix for a neutrino at two different times (140 km/c in blue and 150 km/c in red). It is not possible to see the red lines because they are perfectly behind the blue lines, implying that the flavor evolution with respect to time has stopped. The numerical simulation was carried out using the time-dependent formalism in normal mass ordering and assuming $\Delta \theta = 1^{\circ}$. {\it Right:} The same as the left panel, but for antineutrinos.}
        \label{offdiag}
\end{figure}

Curiously, one would assume that the difference between the results obtained using the time-dependent and the time-independent formalisms implies that the steady-state is not reached in the time-dependent formalism. However, as seen in Fig.~\ref{offdiag}, the off-diagonal components of the density matrix become independent of time when the survival probability ceases to evolve. Figure~\ref{offdiag} shows the real and imaginary parts of the off-diagonal components of $\rho_{ex}$ and $\bar{\rho}_{ex}$ after survival probability stops evolving. It can be seen that the off-diagonal components plotted at two different times overlap perfectly, implying that the steady state has been reached.  

\begin{figure}
\includegraphics[width=0.99\textwidth]{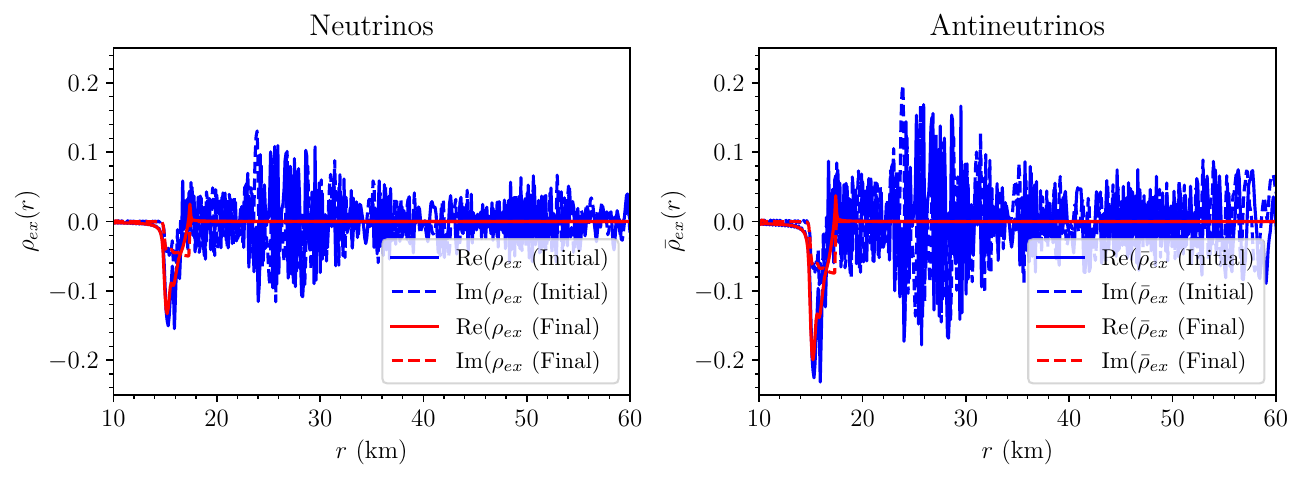}
\centering
\includegraphics[width=0.49\textwidth]{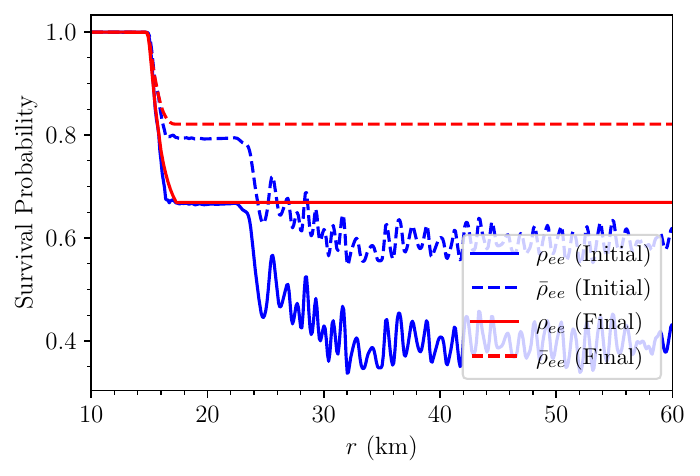}
\caption{{\it Top:} The blue lines show the initial spatial dependence of the real (solid) and imaginary (dashed) parts of the off-diagonal components of the density matrix for neutrinos (left) and antineutrinos (right). The spatial dependence used for the initial state is obtained from the solution of the time-independent formalism. The red lines show the real and imaginary parts of the final state obtained using the time-dependent formalism. Note that red lines are identical to those in Fig.~\ref{offdiag}, implying that the solution is not sensitive to the initial conditions used in the time-dependent formalism. The numerical calculation was performed assuming $\Delta \theta = 1^{\circ}$ and normal mass ordering with 100 angular bins. {\it Bottom:} The same as the top panels, but shows the evolution of the diagonal components of the density matrix. The solid lines are used to denote neutrinos, whereas the dashed lines are used to denote antineutrinos.}
\label{nontrivialini}
\end{figure}

We conduct a test to determine whether the difference in results obtained using the two formalisms is due to the possibility of two different steady states. We first perform the simulation for $\Delta \theta = 1^{\circ}$ and normal mass ordering using the time-independent formalism. We then use the solution thus obtained as the initial condition for the time-dependent formalism. In the case of two possible steady-state solutions, the outputs should depend on the initial conditions used. If the solution obtained using the time-independent formalism is a solution, the time-dependent formalism should not evolve it any further. However, as seen in Fig.~\ref{nontrivialini}, we indeed see an evolution of the neutrino flavor and the neutrino flavor configuration stabilizes to the same configuration that was found in the time-dependent formalism with a trivial radial dependence of neutrino flavor. 

This intriguing outcome leads us to speculate about two different reasons why the time-dependent and time-independent simulations yield different results. One possibility is that although we see the neutrino flavor configuration reach a steady state (see Fig.~\ref{offdiag}), the inevitable but tiny evolution that persists is physical rather than purely numerical. Hence, although the neutrino flavor evolution suggests that a steady state is reached in the case of MNR, that is not entirely true. A second possibility is that the neutrino flavor reaches a steady state, and there are at least two possible solutions. However, one of the solutions obtained using the time-independent formalism is very unstable under time evolution, and a small deviation can cause the system to deviate significantly from the observed solution. It should be noted that if the solution from the time-independent formalism is used as the initial condition in the time-dependent formalism, the flavor configuration obtained is the same as the ones obtained using the time-dependent formalism with trivial initial conditions. This can be clearly seen in Fig.~\ref{nontrivialini}. One can conclude from Fig.~\ref{nontrivialini} that the results obtained using the time-independent formalism are not reliable and are inherently unstable. 

\section{Discussions and Conclusions}
\label{conclusions}
All previous studies of the matter–neutrino resonance (MNR) have assumed that the system evolves toward a steady state. In this work, we presented the first multi-angle numerical study of neutrino flavor evolution in MNR without imposing this assumption. A direct comparison with the results obtained under the steady-state (time-independent) formalism reveals clear and systematic differences. We performed simulations for both mass orderings and found that the steady-state assumption is not justified in either case.

Although previous studies of supernova neutrinos have shown that time-dependent evolution may fail to reach a steady state~\cite{Shalgar:2025ekb}, our results demonstrate that even when a stationary configuration does emerge, as in MNR, the quantitative errors introduced by the steady-state approximation can be substantial. In spherical geometries—such as those explored in Refs.~\cite{Shalgar:2022rjj,Shalgar:2022lvv,Nagakura:2022qko,Xiong:2024tac,Shalgar:2024gjt}, steady-state calculations are particularly problematic because both inward- and outward-propagating neutrino fluxes coexist. The steady-state formalism is therefore most commonly employed in the neutrino-bulb model, where inward fluxes are absent. However, fully time-dependent calculations within the bulb framework remain numerically demanding.

The MNR configuration with a finite opening angle considered here offers a simple and reproducible test system for evaluating the validity of the steady-state approximation. Our results show that the time-independent formalism not only produces quantitatively incorrect survival probabilities but also generates artificial small-scale angular and spatial structures that are absent in the time-dependent treatment. These features are artifacts of the assumption of stationarity rather than physical consequences of flavor dynamics.

This finding has broader implications. The time-independent formalism underlies many early studies of collective neutrino oscillations, including those based on the neutrino-bulb model, which has shaped much of our current understanding of flavor evolution in dense astrophysical environments. Our results indicate that the physical insights drawn from such models must be re-examined in light of their reliance on an intrinsically unstable steady-state approximation.

Future work should extend these analyses to more realistic geometries, such as accretion disks around neutron star mergers, where the interplay of geometry, temporal evolution, and neutrino–neutrino interactions can further enrich the dynamics. Exploring these systems within a fully time-dependent framework will be crucial for achieving a complete and physically consistent description of neutrino flavor evolution in dense astrophysical media.

\section{Acknowledgement}
We thank Markus Ahlers for valuable discussions. We want to thank Marie Cornelius, Mariam Gogilashvili, Manuel Goimil, Maryna Mesiura, and Anna Suliga for going through the manuscript and giving useful comments.
The Tycho supercomputer hosted at the SCIENCE HPC Center at the University
of Copenhagen was used to support the numerical simulations presented in this work.

\appendix

\section{Convergence tests}
\label{AppendixA}
In this Appendix, we justify the numerical convergence of the results presented in the main text. In the context of time-independent numerical simulations, we provide evidence for convergence with respect to the number of angular bins and the tolerance that governs the step size. Additionally, we consider convergence with respect to spatial resolution for the case of time-dependent simulations.

\begin{figure}
\includegraphics[width=0.49\textwidth]{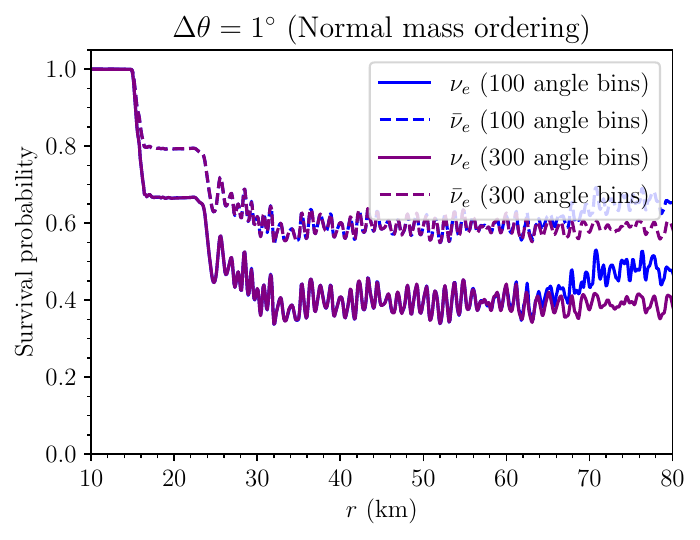}
\includegraphics[width=0.49\textwidth]{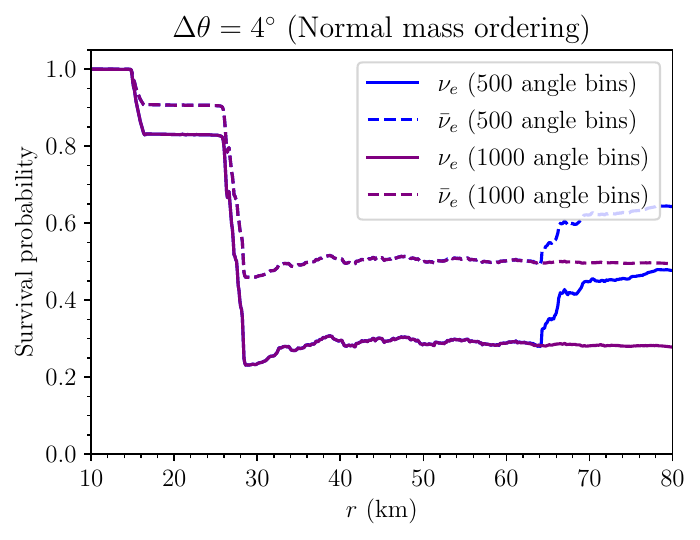}
		\caption{{\it Left: }Survival probability of electron neutrinos and electron antineutrinos as a function of $r$, with $\Delta \theta = 1^{\circ}$ and normal mass ordering calculated using the time-independent formalism. The survival probability has been calculated using 100 (blue) and 300 (purple) angular bins in the time-independent formalism. The two evolve identically up to $\approx 60$ km, after which they diverge from each other. This implies that 100 angular bins are sufficient if the system is evolved up to 60 km. {\it Right: }Same as the left pane,l but for $\Delta \theta = 4^{\circ}$ and the number of angular bins has been changed to 500 (blue) and 1000 (purple).}
		\label{angconv}
\end{figure}

In the case of time-independent simulations, the number of angular bins required is larger than in the time-dependent case. The reason for this is clear from the plot of angular distribution in the main text (see Fig.~\ref{angdist}). The number of angular bins required for convergence depends on the opening angle, $\Delta \theta$, and the spatial range or the time for which the system is evolved. Figure \ref{angconv} shows the evolution of the angle-averaged survival probability for neutrinos and anti-neutrinos for two different numbers of angular bins. In the left panel, one can see that 100 angular bins are sufficient to accurately capture the flavor evolution up to $r \approx 60$ km for the case of $\Delta \theta = 1^{\circ}$. On the other hand, for $\Delta \theta = 4^{\circ}$, one can see that at least 500 angular bins are required for convergence up to $r \approx 60$ km. One can see that although Fig.~\ref{angconv} only shows the results for normal mass ordering, we have verified that the number of angular bins required for convergence is almost identical for inverted mass ordering. 

\begin{figure}
\includegraphics[width=0.49\textwidth]{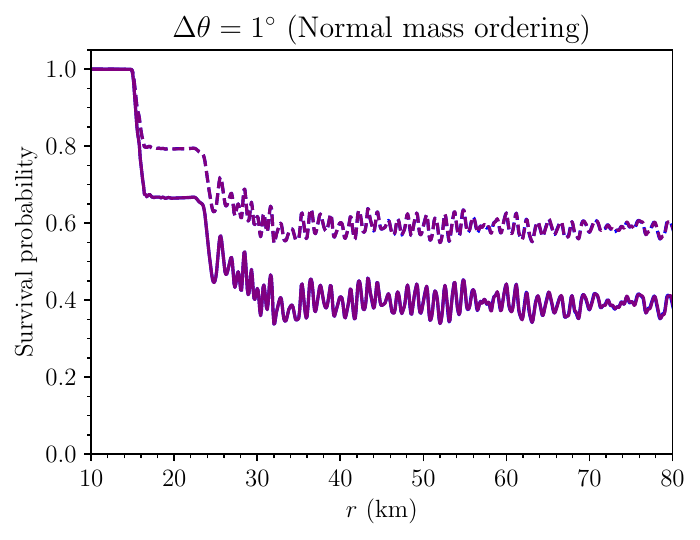}
\includegraphics[width=0.49\textwidth]{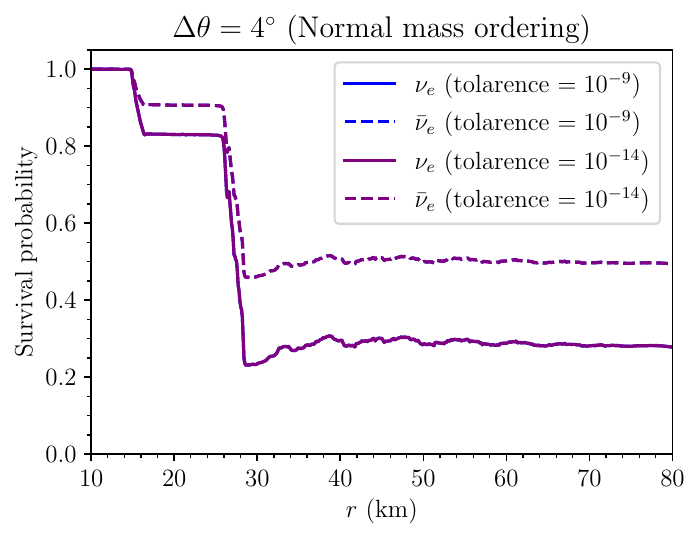}
		\caption{{\it Left: } Survival probability calculated using the time-independent formalism for $\Delta \theta = 1^{\circ}$ and normal mass ordering calculated using absolute and relative tolerance of $10^{-9}$ in blue and $10^{-14}$ in purple. The solid lines depict the survival probability for neutrinos, whereas the dashed lines depict the survival probability for antineutrinos. It is not possible to see the blue lines because they are perfectly behind the purple lines, implying that an absolute and relative tolerance of $10^{-9}$ is more than sufficient. The calculation is done with 300 angular bins. {\it Right: } Same as the left panel but for $\Delta \theta = 4^{\circ}$ and the calculation is done with 1000 angular bins.}
        \label{tolconv}
\end{figure}

In Fig.~\ref{tolconv}, we use the same systems as in Fig.~\ref{angconv} to demonstrate that the choice of our tolerance values is adequate. We perform the flavor evolution calculation with absolute and relative tolerances set to $10^{-9}$, the values we use as default throughout the paper, and compare the results with those obtained using absolute and relative tolerances set to $10^{-14}$. It can be clearly seen in Fig.~\ref{tolconv} that not only a value of $10^{-9}$ sufficient for tolerance but is most likely a conservative choice -- a much larger value of tolerance would have been adequate as the there is virtually no difference in the results -- the plots corresponding to tolerance of $10^{-9}$ is not visible as it is perfectly behind the ones with the tolerance of $10^{-14}$. 

\begin{figure}
\includegraphics[width=0.49\textwidth]{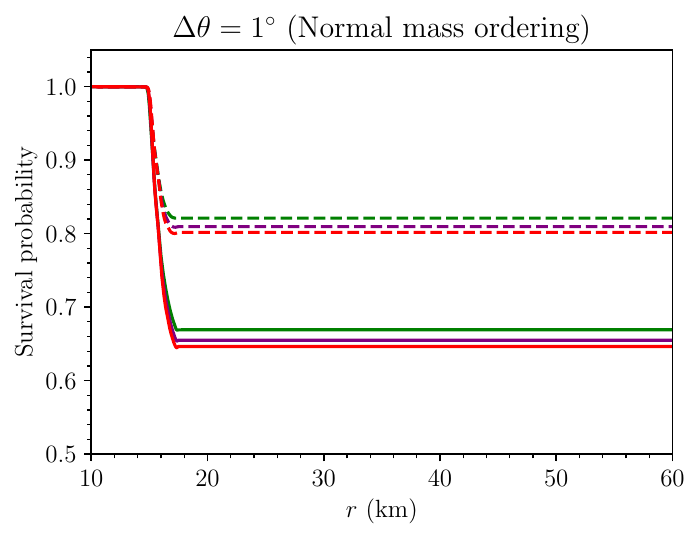}
\includegraphics[width=0.49\textwidth]{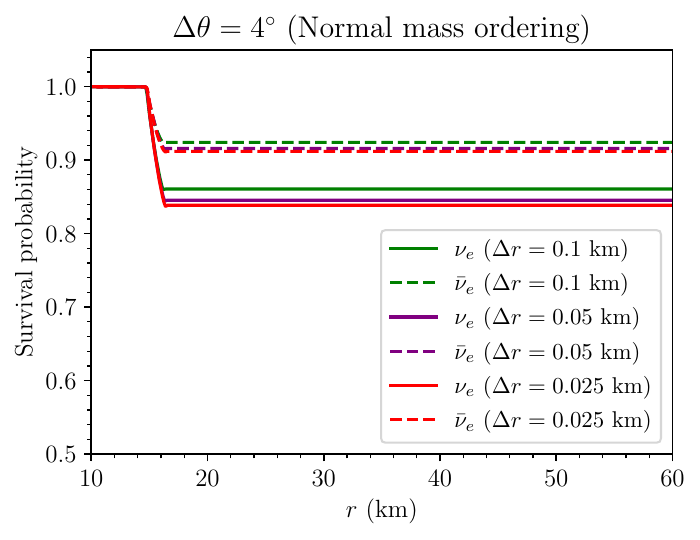}
		\caption{{\it Left: }Survival probability calculated using the time-dependent formalism with three different spatial resolutions. The green, purple, and red lines show the results obtained using $\Delta r = 0.1,0.05$, and $0.025$ km, respectively. The solid and dashed lines depict the survival probabilities for neutrinos and antineutrinos, respectively. The numerical simulation is done assuming $\Delta \theta = 1^{\circ}$ and normal mass ordering. {\it Right: } Same as the left panel but with $\Delta \theta = 4^{\circ}$.}
        \label{radconv}
\end{figure}

In the case of time-dependent formalism, one needs to ensure that a sufficient number of spatial bins are used. In Fig.~\ref{radconv}, we show the angle averaged flavor evolution for $\Delta r = 0.025, 0.05$, and $0.1$ km for representative cases of $\Delta \theta = 1^{\circ}$ and $4^{\circ}$. The results show that the survival probabilities approach an asymptotic value with increased spatial resolution, and that $\Delta r = 0.025$ km is a suitable resolution choice for the cases considered in this paper. It should, however, be noted that we have employed a uniform binning over the spatial range, and it is evident that the dependence of the results on the resolution primarily stems from a narrow spatial region with a dip in the survival probability. It is thus possible to perform a more efficient calculation by using a nonuniform grid, but we do not explore that topic in this paper. 

\begin{figure}
\includegraphics[width=0.99\textwidth]{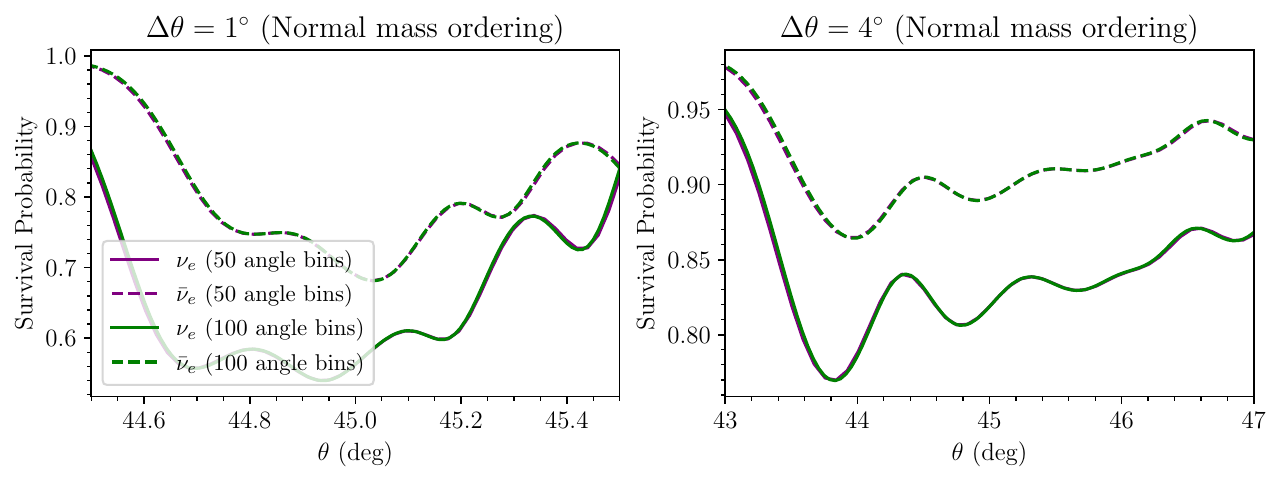}
		\caption{{\it Left:} Angular distribution of neutrinos at 60 km calculated in the time-dependent formalism and $\Delta\theta = 1^{\circ}$ calculated using 50 (purple) angular bins and 100 (green) angular bins. The purple lines are not visible as they are perfectly behind the green line. {\it Right:} Same as the left panel but for $\Delta \theta = 4^{\circ}$.}
        \label{td-radconv}
\end{figure}

As already alluded to in the main text, the number of angular bins required for convergence is very different in the case of time-independent and time-dependent calculations. In the case of time-dependent calculations, we observe that convergence is achieved with a relatively small number of angular bins. This is shown in Fig.~\ref{td-radconv}. 

In summary, the convergence tests presented here confirm that the numerical results discussed in the main text are robust with respect to angular resolution, spatial resolution, and solver tolerances.

\bibliographystyle{ws-ijmpd}
\bibliography{main.bib}

\end{document}